%% file: supplementary_materials.tex
\pdfoutput=1
\documentclass[12pt,a4paper]{article}
\title{
Supplementary Materials for \\ \textbf{LV Barcoding: locality sensitive hashing-based tool for fast species identification based on DNA barcoding}
}
\author{Long Fan$^1$ and Ka Hou Chu$^{1,*}$}
\date{}
\usepackage{pdfpages}%
\usepackage{graphicx}
\usepackage{latexsym}
\usepackage{amsmath}
\usepackage{amssymb}

\begin{document}


# LV Barcoding: locality sensitive hashing-based tool for rapid species identification in DNA barcoding

Long Fan[1] and Ka Hou Chu[1,*]

[1] School of Life Sciences, The Chinese University of Hong Kong, Shatin, Hong Kong, China.

**ABSTRACT**

DNA barcoding has emerged as a cost-effective approach for species identification. However, the scarcity of tools used for searching the booming reference database becomes an obstacle, currently with BLAST as the only practical choice.

Here, we propose a program - LV Barcoding - based on both the random hyperplane projection-based locality sensitive hashing method and the composition vector-based VIP Barcoding for fast species identification. The performance of LV Barcoding is assessed on the data release of BOLD. LV Barcoding has higher accuracy than BLAST, and is able to match a single query against ~114,000 reference barcodes within 10 seconds on a desktop computer.

**Availability**: The program is available at http://msl.sls.cuhk.edu.hk/vipbarcoding/
**Contact:** kahouchu@cuhk.edu.hk

## 1 INTRODUCTION

DNA barcoding uses short gene regions as internal species tags, and assigns individuals to given species according to their sequences at the standardized locus (Hebert, et al., 2003). The first phase of the International Barcode of Life Project plans to barcode five million specimens. At present, the Barcode of Life Data Systems (BOLD) (Ratnasingham and Hebert, 2007) has 150,610 COI barcodes in its latest data release 4.75 - v1 (Mar-31-2014). However, searching such a large and booming database makes the researchers committed to use BLAST (Altschul, et al., 1990), although the accuracy of BLAST has been criticized as compared with the other alternatives as these are much slow in data processing (Austerlitz, et al., 2009; Koski and Golding, 2001; Meyer and Paulay, 2005).

To develop a rapid software for DNA barcoding, we recently presented the composition vector-based VIP Barcoding (Fan, et al., 2014), in which a two-stage hybrid algorithm was adopted. First, a cosine similarity-based alignment-free method is utilized to screen the reference database in reducing the searching space. The alignment-based K2P distance nearest neighbor method is then employed to analyze the smaller dataset after screening at the first stage. Yet the time-consuming brute force searching of the first step poses a challenge for larger scalability.

Here, we propose **L**ocality sensitive hashing-based **V**IP **Barcoding** (LV Barcoding) - a three-stage hybrid algorithm by integrating random hyperplane projection-based locality sensitive hashing (RHP) (Charikar, 2002) with VIP Barcoding. LV Barcoding runs faster than VIP Barcoding without sacrificing accuracy, and both of them positively identified more queries than BLAST. Therefore, LV Barcoding could serve as an excellent alternative to BLAST in rapid species identification.

## 2 METHODS

### 2.1 Workflow of LV Barcoding

RHP is designed for the approximate evaluation of cosine similarity. Since both RHP and the first step of VIP Barcoding use cosine similarity to estimate the similarity of two sequences, RHP is optimal to be integrated with VIP Barcoding among the locality sensitive hashing methods. Therefore RHP works as a rough and fast screening, followed by the first step of VIP Barcoding performing as a relatively more precise second screening.

RHP is implemented according to Charikar (2002). The details are described in supplementary materials. Briefly, LV Barcoding employs RHP to segment the reference database into smaller clusters (i.e., subspaces or buckets) in the construction of reference database. It could be understood as the step in which a high-dimensional space is split into many smaller subspaces by the hyperplanes of RHP, and each barcode was projected in a certain subspace. Since RHP could rapidly estimate cosine similarity, similar barcodes are projected into the same subspace. In a query search, LV Barcoding projects the query into one of subspaces. Finally, LV Barcoding uses the two-stage hybrid algorithm of VIP Barcoding (Fan, et al., 2014), to match the query against the reference barcodes co-localized with the query in the same subspace.

### 2.2 Implementation

LV Barcoding written in C++ can run on Windows, Mac and Linux operating systems. By utilizing Qt library, LV Barcoding not only provides a user-friendly graphical user interface, but also performs parallel computation using multi-core processor. Boost library is used for matrix operations of RHP.

## 2.3 Comparison of LV Barcoding with other methods

The compared methods comprise BLAST, VIP Barcoding, and LV Barcoding. We downloaded COI dataset of Data Package Release 4.50 - v1 of BOLD for the tests. After the redundantly identical barcodes from a single species were deleted, there were 114,049 unique records in the test dataset. To test the effect of the size of the dataset, another five artificial datasets were constructed by randomly choosing 20,000, 40,000, 60,000, 80,000, and 100,000 barcodes from the test dataset, respectively. From each dataset, 100 sequences were randomly sampled and used as queries in a single trial. This trial was repeated 10 times, each with 100 new independent queries. Moreover, the leave-one-out cross-validation (for details, see Fan, et al., 2014) was adopted to evaluate the accuracy at the species level. All experiments were carried out on a desktop computer equipped with Intel i7-3770 CPU, 16 GB RAM and the Windows 7 64-bit system.

## 3 RESULTS

### 3.1 Accuracy Benchmark

The accuracy comparison using a total of 1,000 random queries is summarized in Table 1. Although we just randomly extracted a small portion of the complete dataset as queries, it is apparent that LV Barcoding and VIP Barcoding are similarly accurate, which are consistently higher than Blastn's.

**Table 1.** Numbers of positive identification at the species level for 1,000 queries. The definition of positive identification is cited from Fan, et al., (2014).

| Dataset | Barcodes | Blastn* | LV Barcoding | VIP Barcoding |
|---|---|---|---|---|
| 1 | 20,000 | 614 | 619 | 619 |
| 2 | 40,000 | 675 | 683 | 683 |
| 3 | 60,000 | 762 | 774 | 774 |
| 4 | 80,000 | 775 | 780 | 780 |
| 5 | 100,000 | 810 | 819 | 819 |
| 6 | 114,049 | 809 | 832 | 832 |

* Blastn is included in NCBI BLAST 2.2.29+.

### 3.2 Speed Benchmark

LV Barcoding on the average spent approximate 6.5 seconds in completing a single query against 114,049 reference barcodes. As shown in Figure 1, the extent of increase (slope) in the time cost of LV Barcoding with the increase in dataset size apparently decreases as compared with VIP Barcoding. With both the accuracy and speed taken into consideration, we regard LV Barcoding superior to Blastn for DNA barcoding.

## ACKNOWLEDGMENTS

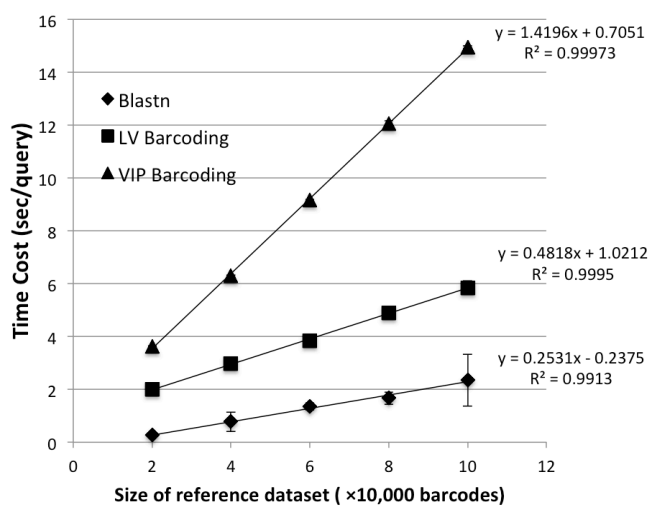

Figure 1. Linear regression of time costs of three different methods. Error bars represent standard deviations of ten replicates of 100 searches. Most error bars are within the size of the symbols. The maximum number of threads of the computer was used for Blastn for making the comparison impartial.

%
\setcounter{page}{1}
\maketitle
\noindent $^1$School of Life Sciences, The Chinese University of Hong Kong, Shatin, Hong Kong, China.
\newline
\newline
\newline
\indent This file describes the algorithm of LV Barcoding in detail. LV Barcoding integrates random hyperplane projection-based locality sensitive hashing (RHP) (Charikar, 2002) and VIP Barcoding (Fan, et al., 2014), and its workflow is illustrated in Figure~\ref{figure:1}. The details of RHP in LV Barcoding is shown in Figure~\ref{figure:2}. At the beginning, a list of hash functions ($h_{r}$, i.e., hyperplanes) is randomly generated for each reference database according to Charikar (2002), and then each reference barcode ($R_{i}$) is projected into different subspaces (i.e., buckets indexed using bit vectors and shown in the black rectangle) by RHP. After the projection, similar reference barcodes exists in the same bucket. When a query ($Q$) is input, the same list of hash functions would be used to project the query into a certain bucket shown in red rectangle. Then the two-stage hybrid algorithm of VIP Barcoding would be used to match a query against the reference barcodes co-localized with the query in the same subspace. Because the number ($n$) of the barcodes at a single bucket would be much smaller than the total size ($N$) of the reference database, the running time for searching would be greatly reduced in LV Barcoding as compared to VIP Barcoding.
\input{./figure/1}
\vspace{4cm}  
\input{./figure/2}
\newpage
\section{Projection of reference barcodes}
\indent Firstly, each reference barcode would be represented using its composition vector constructed according to Fan et al., (2014), but the length of the k-mers of the composition vector used in this step is independent to the composition vector utilized in the two-stage hybrid algorithm of VIP Barcoding. Secondly, RHP generated a series of random hyperplanes as hash functions $h_{\vec{r}}(x)$ for each reference database. Let $D$ equal to the number of the random vectors, so \[h_{\vec{r}}(x)=<h_{\vec{r}_1}(x),h_{\vec{r}_2}(x),\cdots, h_{\vec{r}_D}(x)>.\] Each hyperplane could be briefly represented by its normal vector [i.e., a random vector ($\vec{r_i}$) of which each element is randomly generated from a Gaussian distribution $N(0,1)$]. Finally, RHP utilizes these functions to encode the composition vector as a fixed-size bit vector (i.e., bucket index), and the length of the bit vector equals to the number ($D$) of the random vectors. The hashing step is such that, the composition vector ($\vec{c}$) of each sequence could be mapped into a certain bucket indexed by a $D$-bits binary vector ($B$), of which the $i_{th}$ element is calculated through the dot product between $\vec{c}$ and the $i_{th}$ random vector by Equation~\ref{equation:1}:
	\begin{equation}
	\label{equation:1}
	B_{\vec{c}}[i]=h_{\vec{r_i}}(\vec{c}) = \left\{\ 
	\begin{matrix}
	 1 & if\ \vec{r_i}\cdot \vec{c} \geq 0 \\ 
	 0 & if\ \vec{r_i}\cdot \vec{c}  < 0 \\
	\end{matrix} \right.
	\phantom{.}.
	\end{equation}
Therefore, $B$ is given by \smash{$B_{\vec{c}}=<h_{\vec{r}_1}(\vec{c}),h_{\vec{r}_2}(\vec{c}),\cdots, h_{\vec{r}_D}(\vec{c})>$}.
\newline
\subsection{Approximate cosine similarity estimated using RHP}
\indent Eq.~\ref{equation:1} has a simple geometric interpretation. To illustrate this clearly, let us assume dimensionality of composition vector equaling to 2, and then the hyperplanes could be simplified to be some random lines of two-dimensional plane. Here, we utilizes two lines cross the origin to estimate which vector from $\vec{c_2}$, $\vec{c_3}$ and $\vec{c_4}$ has the largest cosine similarity with $\vec{c_1}$. Figure~\ref{figure:3} shows that each line divides the plane into two sides. $\vec{c_1}$ and $\vec{c_2}$ are located at the same sides of both Line 1 and Line 2, while $\vec{c_1}$ and $\vec{c_3}$ are separated by Line 1, and $\vec{c_1}$ and $\vec{c_4}$ are separated by both Line 1 and Line 2. So it is concluded that $\vec{c_2}$ is more similar to $\vec{c_1}$. Meanwhile, it could be noticed that whether two vectors exist at the same side of a certain line is a binary result (i.e., 0 is false and 1 is true or otherwise). Therefore, we can directly calculate the hamming distance of two bit vectors (i.e., bucket indexes) generated by RHP, and use this hamming distance to approximately evaluate the cosine similarity of original two composition vectors according to the following relationship Equation (Charikar, 2002):
	\begin{equation}
	\label{equation:2}
	cos(\theta(\vec{u},\vec{v}))=cos(\frac{\pi\cdot Hamming(B_{\vec{u}},B_{\vec{v}})}{D}) \phantom{,},
	\end{equation}
where $\theta(,)$ denotes the angle between two vectors, and $Hamming(,)$ is the hamming distance of two bit vectors.
\input{./figure/3}
\indent Similarly, the intuition behind Eq.~\ref{equation:1} and Eq.~\ref{equation:2} is that, if two composition vectors evaluated through cosine similarity are similar to each other, then with high probability the random hyperplane projections will enable them to be located at the same sides of these hyperplanes. On the other hand, two vectors separated by a lager included angle are very likely to be projected into different sides of hyperplanes.

\subsection{Accuracy ensured under adjustable probability}
\indent For any two vectors $\vec{u}$ and $\vec{v}$, the probability that a single random projection collides (i.e., two vectors at the same side of one random hyperplane) is:
	\begin{equation}
	\label{equation:3}
	Pr[h_r(\vec{u})=h_r(\vec{v})] = 1 -\frac{\theta (\vec{u},\vec{v})}{\pi } \phantom{.}.
	\end{equation}
Given \smash{$cos(\theta(\vec{u},\vec{v}))=t$}, then \smash{$Pr[h_{\vec{r}}(\vec{u})=h_{\vec{r}}(\vec{v})] = 1 -\frac{cos^{-1}(t)}{\pi }$}. Since we have $D$ hash functions, it infers:
	\begin{equation}
	\label{equation:4}
	Pr[B_{\vec{u}}=B_{\vec{v}}] = (1 -\frac{cos^{-1}(t)}{\pi })^D \phantom{.}.
	\end{equation}
\indent Naturally, false positives (FP) and false negatives (FN) are the common problems of all LSH-based approaches. More projections (i.e., larger $D$) would help to reduce the number of comparisons which were required in the following stage. In another word, the percentage of FP in all positives would decrease. However, increasing $D$ will also accumulate error of each projection, i.e., the number of FNs increases gradually. For instance, two sequences $S_x$ and $S_y$ are similar but not exactly same, then a FN (i.e., to project two sequences into different buckets) tends to occur especially when larger $D$ is used. To alleviate this issue, we can repeat the whole procedure multiple times, and FNs can be substantially reduced by iterating the courses and using differently generated \smash{$h_{\vec{r}}$}. Consequently, the probability of successfully projecting a really similar pair of composition vectors into the same bucket in at least one trial of $M$ times meets the following equation:
	\begin{equation}
	\label{equation:5}
	\begin{aligned}
	&Pr[\vec{u},\vec{v}\ in\ same\ bucket\ in\ \geqslant 1\ trial\ of\ M\ times\ |\ cos(\theta (\vec{u},\vec{v}))=t ]\\
	=&Pr[B_{\vec{u}}=B_{\vec{v}}\ in\ \geqslant 1\ trial\ of\ M\ times\ |\ cos(\theta (\vec{u},\vec{v}))=t ]\\
	=&1-[1-(1-\frac{cos^{-1}(t)}{\pi})^D]^M
	\end{aligned}
	\end{equation}
\newline
\indent Suppose we set a threshold ($t$) for cosine similarity and expect that all the reference barcodes within this threshold could be accurately put into the same bucket of the query in at least one trial of $M$ times, the probability of success should be calculated according to following three steps.
\newline\newline
(1) In terms of Eq.~\ref{equation:2}, we can count the Hamming distance threshold ($T$) given by:
	\begin{equation}
	\label{equation:6}
	T=\frac{D\cdot cos^{-1}(t)}{\pi} \phantom{.}.
	\end{equation}
\newline\newline
(2) Using Eq.~\ref{equation:4} and Eq.~\ref{equation:6} as reference, we can infer the following relationship:
	\begin{equation}
	\label{equation:7}
	Pr[Hamming(B_{\vec{u}},B_{\vec{v}}) \leqslant T\ |\ cos(\theta (\vec{u},\vec{v}))=t ]= \sum_{i=0}^{T} \binom{D}{i} P^i (1-P)^{D-i} \phantom{,},
	\end{equation}
where \smash{$P=cos^{-1}(t)/\pi=T/D$}.
\newline\newline
(3) Using Eq.~\ref{equation:5}, Eq.~\ref{equation:6} and Eq.~\ref{equation:7} as reference, we can calculate the probability that all the reference barcodes within this threshold could be accurately put into the same buckets of the query in at least one trial of $M$ times like this:
	\begin{equation}
	\label{equation:8}
	\begin{aligned}
	&Pr[\vec{u},\vec{v}\ in\ same\ bucket\ in\ \geqslant 1\ trial\ of\ M\ times\ |\ Hamming(B_{\vec{u}},B_{\vec{v}}) \leqslant T]\\
	=&1-[1-\sum_{i=0}^{T} \binom{D}{i} P^i (1-P)^{D-i}]^M \phantom{,},
	\end{aligned}
	\end{equation}
where \smash{$P=cos^{-1}(t)/\pi=T/D$}.
And it could be noticed that Eq.~\ref{equation:8} and Eq.~\ref{equation:5} are the same, given $T=0$, and by adjusting $M$, $t$, and $D$, we can modify the probability of Eq.~\ref{equation:8}.

\section{Parameters of LV Barcoding and their effects}
\indent In this subsection, we describe the parameters of LV Barcoding algorithm in details.
\subsection{Length of k-mer ($w$)}
\indent As we mentioned above, the length $w$ of k-mer used in RHP is independent on the parameter $k$ of VIP Barcoding (for details, see Fan et al., (2014)). The use of a long $w$ results in the fast growth of the dimension of the composition vector in RPH. Since LSH generates the hash functions of which the length of each hash vector is same to $w$, the time used for producing hash vector and calculating dot product increases with the increase of $w$. Meanwhile, it requires more space of hard disk to store those hash functions for the reusage during each time of retrival. More importantly, the use of a long k-mer would prevent RPH from distinguishing similar sequences robustly, such that only almost identical sequences are grouped into a bucket. On the other hand, the use of a short k-mer would decrease the sensitivity in partitioning the reference barcodes, which causes very slight improvement in speed. Therefore, we choose $w$ from 4, 5 and 6 in preliminary parameter setting.

\subsection{Threshold of cosine similarity ($t$)}
\indent The parameter $t$ was implemented in order to allow for pairs of strings that are identical or similar to go through the RPH filter and be allocated into the same bucket. The use of $t$ value that is too small or too large would reduce the accuracy and speed, respectively. Hence, $t$ is chosen from 0.8 to 0.9 in preliminary parameter setting.

\subsection{Number of hash functions or the length of bucket index in a trial ($D$)}
\indent In contrast to $t$, the use of $D$ value that is too small or too large would decrease the speed and accuracy, respectively, since the use of a small $D$ would enable the sequences with many mismatches to be projected into the same bucket, while the use of a large $D$ would only put identical sequences into one bucket. Since RHP should not wrongly delete the true positive barcodes, $D$ is selected from 14 to 17 in preliminary parameter setting.

\subsection{Number of multiple trials ($M$)}
\indent The utilization of multiple hash functions within the RPH is helpful for reducing the FNs, i.e., two similar sequences could gain a higher chance of being segmented into the same bucket after several trials. But the adverse effect of enlarged $M$ is the increased number of FPs. More FPs would not affect the accuracy of the final species identification, but slow the speed of the whole process.

\section{Parameter setting}
It is expected in Eq.~\ref{equation:8} that \[Pr[\vec{u},\vec{v}\ in\ same\ bucket\ in\ \geqslant 1\ trial\ of\ M\ times] \geqslant 0.95,\] which enables LV Barcoding to preserve the same degree of accuracy as VIP Barcoding. Under this premise, we tested combinations of different values of the above parameters using COI dataset of Data Package Release 4.50 - v1 of BOLD. Finally, we set the default parameters of LV Barcoding as follows: $w=6,\ t=0.9,\ D=14,\ M=4$. In practice, $t$ is invisible in the panel of parameter setting in LV Barcoding, which provides optional hamming distance threshold $T$ for which the default value is 2.

\section*{References}

\noindent
Charikar, M.S. Similarity estimation techniques from rounding algorithms. Proceedings of the thiry-fourth annual ACM symposium on Theory of computing. Montreal, Quebec, Canada: ACM; 2002. p. 380-388.
\newline
\newline
\noindent
Fan, L., et al. VIP Barcoding: composition vector-based software for rapid species identification based on DNA barcoding. Molecular Ecology Resources 2014; doi: 10.1111/1755-0998.12235.

\end{document}

%% file: figure/1.tex
\begin{figure}[h!]
  \centering
  \includegraphics[trim=0cm 0cm 0cm 0cm, width=0.92\textwidth]{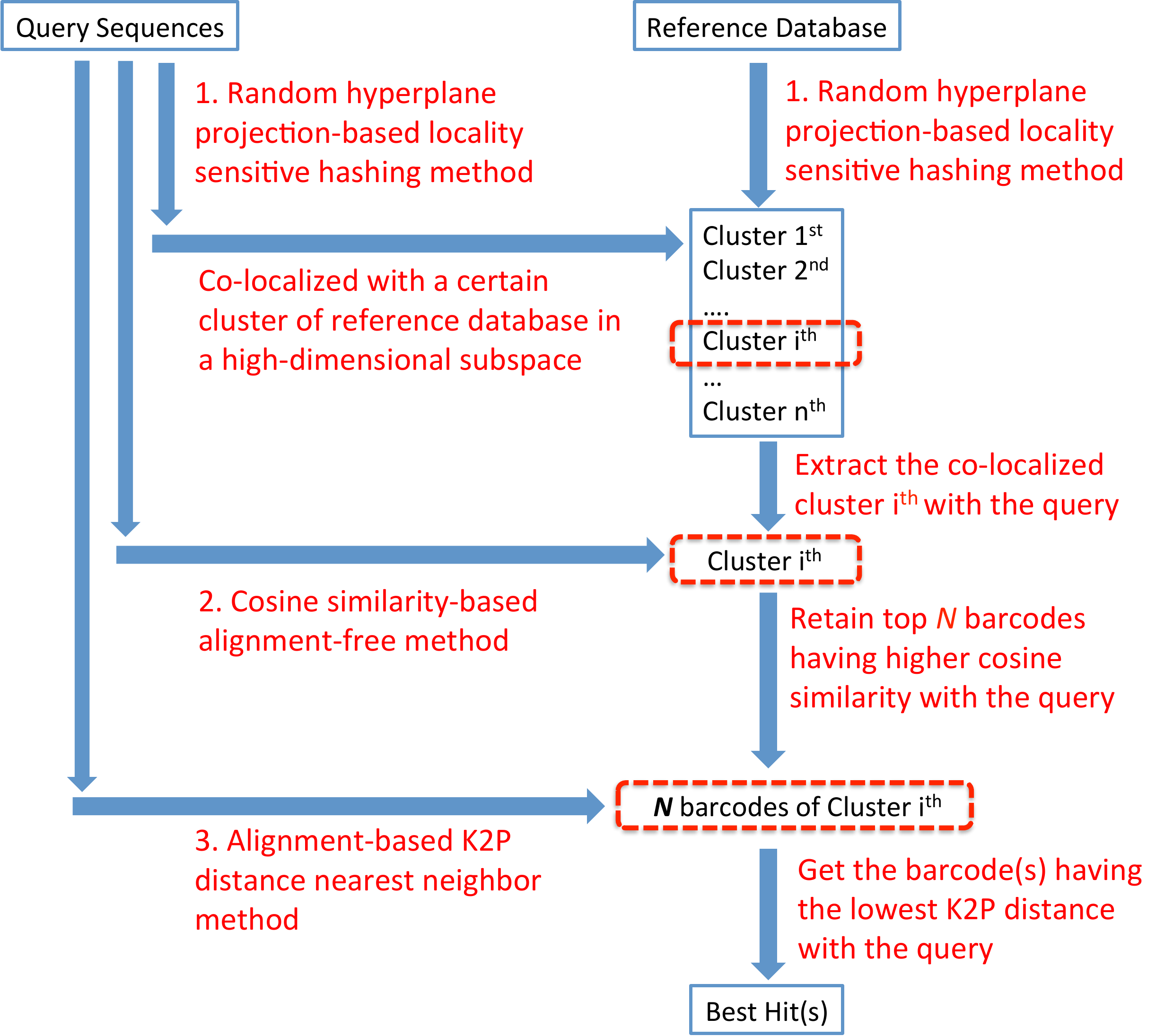}  
  \caption[Workflow of LV Barcoding]{Workflow of LV Barcoding.}\label{figure:1}
\end{figure}

%% file: figure/2.tex
\begin{figure}[h!]
  \centering
  \includegraphics[trim=0cm 0cm 0cm 0cm, width=0.8\textwidth]{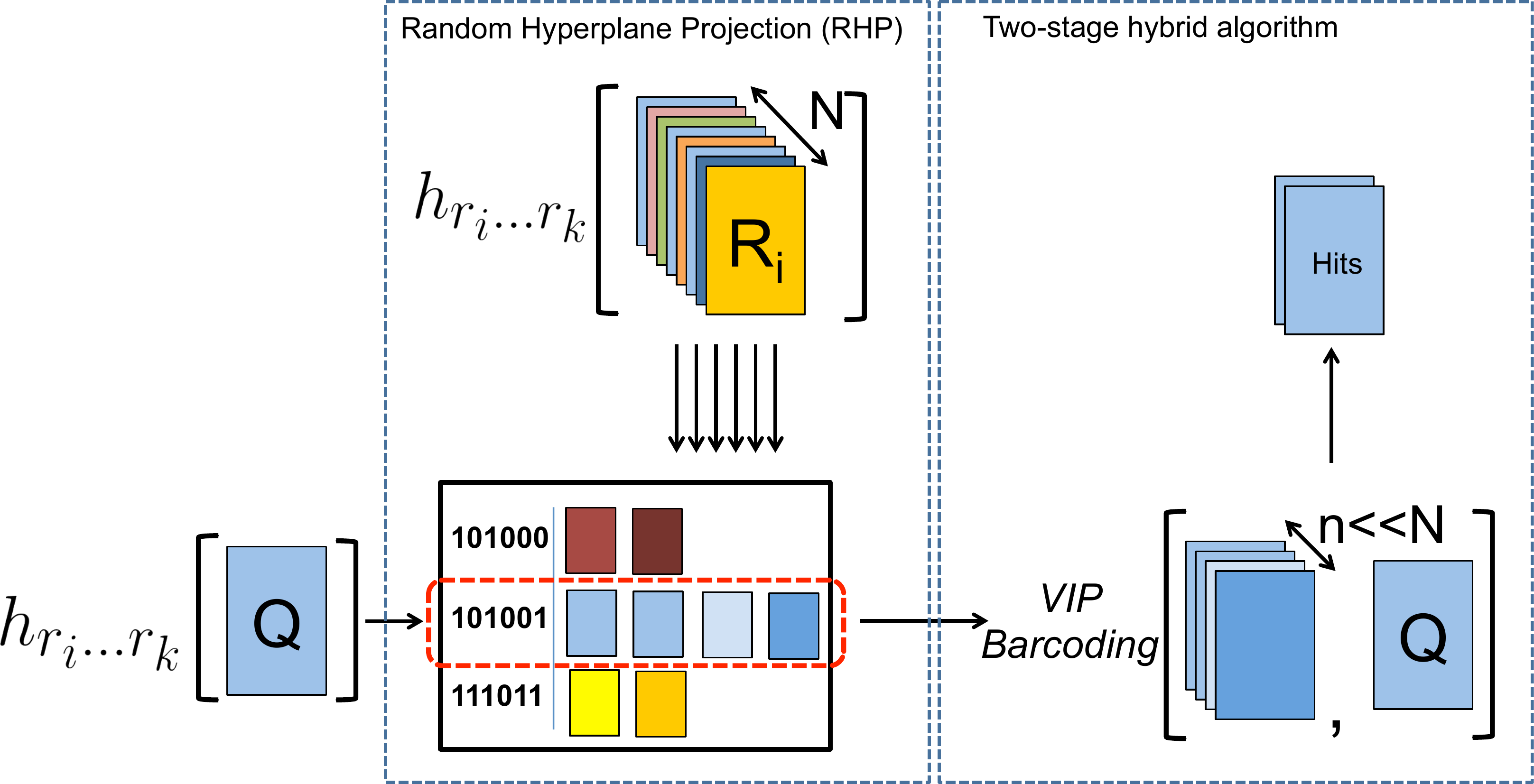}  
  \caption[Details of RHP in LV Barcoding]{Details of RHP in LV Barcoding.}\label{figure:2}
\end{figure}

%% file: figure/3.tex
\begin{figure}
  \centering
  \includegraphics[trim=0cm 0cm 0cm 0cm, width=0.8\textwidth]{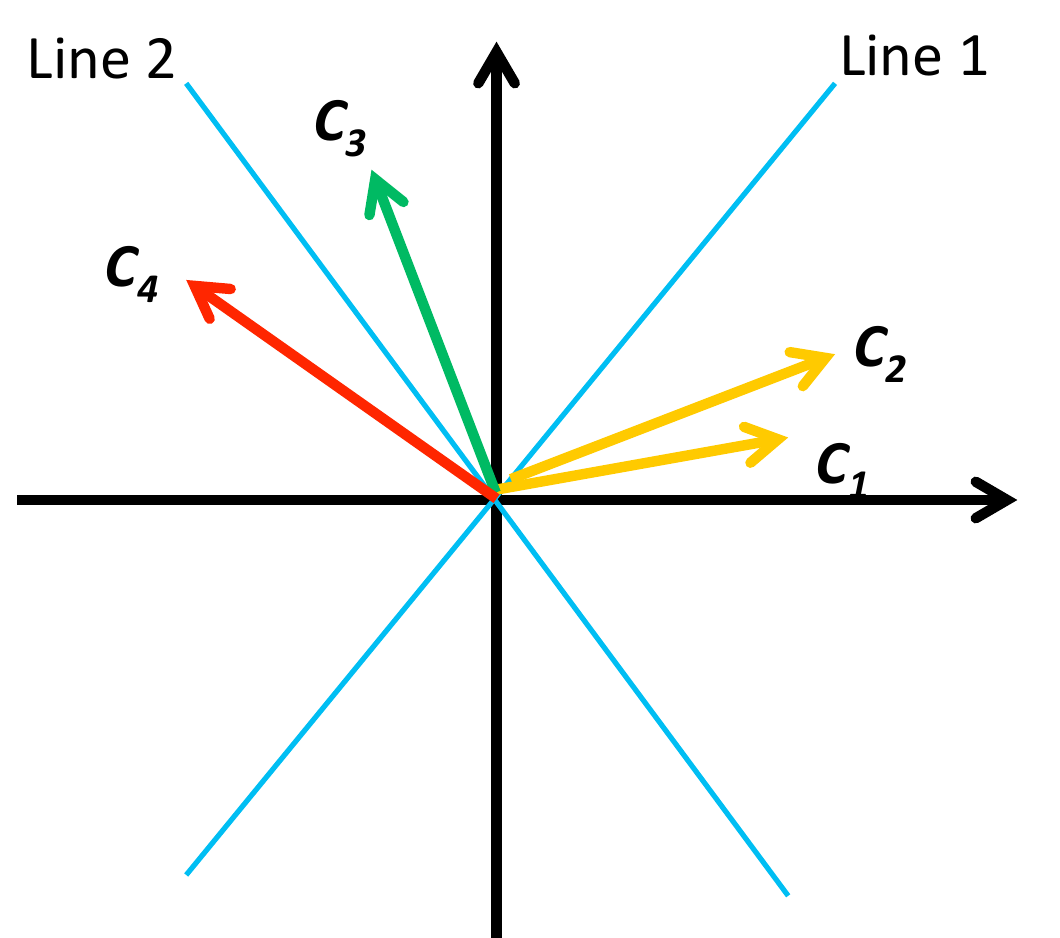}  
  \caption[Illustration of projection of hash function]{Illustration of projection of hash function}\label{figure:3}
\end{figure}
\newpage